\input harvmac
\input amssym
\input epsf

\def\CE{{\cal E}}
\def\CQ{{\cal Q}}


\lref\AK{
  O.~Aharony and D.~Kutasov,
  ``Holographic Duals of Long Open Strings,''
  Phys.\ Rev.\  D {\bf 78} (2008) 026005
  [arXiv:0803.3547 [hep-th]].
}
\lref\ASY{
  O.~Aharony, J.~Sonnenschein and S.~Yankielowicz,
  ``A holographic model of deconfinement and chiral symmetry restoration,''
  Annals Phys.\  {\bf 322} (2007) 1420
  [arXiv:hep-th/0604161].
}
\lref\BLW{
  A.~Buchel, P.~Langfelder and J.~Walcher,
  ``Does the tachyon matter?,''
  Annals Phys.\  {\bf 302} (2002) 78 
  [arXiv:hep-th/0207235].
}
\lref\BSS{
  O.~Bergman, S.~Seki and J.~Sonnenschein,
  JHEP {\bf 0712} (2007) 037 
  [arXiv:0708.2839 [hep-th]].
}
\lref\BW{
  A.~Buchel and J.~Walcher,
  ``The tachyon does matter,''
  Fortsch.\ Phys.\  {\bf 51} (2003) 885
  [arXiv:hep-th/0212150].
}
\lref\CPS{
  R.~Casero, A.~Paredes and J.~Sonnenschein,
  ``Fundamental matter, meson spectroscopy and non-critical string/gauge
  duality,''
  JHEP {\bf 0601} (2006) 127
  [arXiv:hep-th/0510110].
}
\lref\DN{
  A.~Dhar and P.~Nag,
  ``Tachyon condensation and quark mass in modified Sakai-Sugimoto model,''
  Phys.\ Rev.\  D {\bf 78} (2008) 066021 
  [arXiv:0804.4807 [hep-th]].
}
\lref\DP{
  L.~Da Rold and A.~Pomarol,
  ``Chiral symmetry breaking from five dimensional spaces,''
  Nucl.\ Phys.\  B {\bf 721} (2005) 79 
  [arXiv:hep-ph/0501218].
}
\lref\EKSS{
  J.~Erlich, E.~Katz, D.~T.~Son and M.~A.~Stephanov,
  ``QCD and a Holographic Model of Hadrons,''
  Phys.\ Rev.\ Lett.\  {\bf 95} (2005) 261602
  [arXiv:hep-ph/0501128].
}
\lref\Ga{
  M.~R.~Garousi,
  ``D-brane anti-D-brane effective action and brane interaction in open  string channel,''
  JHEP {\bf 0501} (2005) 029 
  [arXiv:hep-th/0411222].
}
\lref\GOR{
  M.~Gell-Mann, R.~J.~Oakes and B.~Renner,
  ``Behavior of current divergences under SU(3) x SU(3),''
  Phys.\ Rev.\  {\bf 175} (1968) 2195.
}
\lref\GH{
  M.~R.~Garousi and E.~Hatefi,
  ``On Wess-Zumino terms of Brane-Antibrane systems,''
  Nucl.\ Phys.\  B {\bf 800} (2008) 502
  [arXiv:0710.5875 [hep-th]];
  ``More on WZ action of non-BPS branes,''
  JHEP {\bf 0903} (2009) 008
  [arXiv:0812.4216 [hep-th]].
}
\lref\GP{
  D.~Gepner and S.~S.~Pal,
  ``Chiral symmetry breaking and restoration from holography,''
  arXiv:hep-th/0608229.
}
\lref\HHLY{
  K.~Hashimoto, T.~Hirayama, F.~L.~Lin and H.~U.~Yee,
  ``Quark Mass Deformation of Holographic Massless QCD,''
  JHEP {\bf 0807} (2008) 089
  [arXiv:0803.4192 [hep-th]].
}
\lref\HHM{
  K.~Hashimoto, T.~Hirayama and A.~Miwa,
  ``Holographic QCD and pion mass,''
  JHEP {\bf 0706} (2007) 020
  [arXiv:hep-th/0703024].
}
\lref\HN{
  K.~Hashimoto and S.~Nagaoka,
  ``Recombination of intersecting D-branes by local tachyon condensation,''
  JHEP {\bf 0306} (2003) 034
  [arXiv:hep-th/0303204].
}
\lref\JL{
  N.~Jokela and M.~Lippert,
  ``Inhomogeneous tachyon dynamics and the zipper,''
  JHEP {\bf 0908} (2009) 024
  [arXiv:0906.0317 [hep-th]].
}
\lref\KL{
  P.~Kraus and F.~Larsen,
  ``Boundary string field theory of the DD-bar system,''
  Phys.\ Rev.\  D {\bf 63} (2001) 106004 
  [arXiv:hep-th/0012198].
}
\lref\KMM{
  D.~Kutasov, M.~Marino and G.~W.~Moore,
  ``Remarks on tachyon condensation in superstring field theory,''
  arXiv:hep-th/0010108.
}
\lref\KMMW{
  M.~Kruczenski, D.~Mateos, R.~C.~Myers and D.~J.~Winters,
  ``Towards a holographic dual of large-$N_c$ QCD,''
  JHEP {\bf 0405} (2004) 041 
  [arXiv:hep-th/0311270].
}
\lref\KW{
  C.~Kennedy and A.~Wilkins,
  ``Ramond-Ramond couplings on brane-antibrane systems,''
  Phys.\ Lett.\  B {\bf 464} (1999) 206 
  [arXiv:hep-th/9905195].
}
\lref\LLM{
  N.~D.~Lambert, H.~Liu and J.~M.~Maldacena,
  ``Closed strings from decaying D-branes,''
  JHEP {\bf 0703} (2007) 014
  [arXiv:hep-th/0303139].
}
\lref\LP{
  F.~Leblond and A.~W.~Peet,
  ``SD-brane gravity fields and rolling tachyons,''
  JHEP {\bf 0304} (2003) 048 
  [arXiv:hep-th/0303035].
}
\lref\Ma{
  J.~M.~Maldacena,
  ``The large N limit of superconformal field theories and supergravity,''
  Adv.\ Theor.\ Math.\ Phys.\  {\bf 2} (1998) 231 
  [Int.\ J.\ Theor.\ Phys.\  {\bf 38} (1999) 1113]
  [arXiv:hep-th/9711200].
}
\lref\Po{
J.~Polchinski, ``String Theory, Vol.~1: An Introduction to the Bosonic String''; ``String Theory, Vol.~2: Superstring Theory and Beyond,'' Cambridge University Press (1988).
}
\lref\PR{
  E.~Pomoni and L.~Rastelli,
  ``Intersecting Flavor Branes,''
  arXiv:1002.0006 [hep-th].
}
\lref\PSZ{
  K.~Peeters, J.~Sonnenschein and M.~Zamaklar,
  ``Holographic melting and related properties of mesons in a quark gluon
  Phys.\ Rev.\  D {\bf 74} (2006) 106008
  [arXiv:hep-th/0606195].
}
\lref\SS{
  T.~Sakai and S.~Sugimoto,
  ``Low energy hadron physics in holographic QCD,''
  Prog.\ Theor.\ Phys.\  {\bf 113} (2005) 843 
  [arXiv:hep-th/0412141].
}
\lref\SSi{
  T.~Sakai and S.~Sugimoto,
  ``More on a holographic dual of QCD,''
  Prog.\ Theor.\ Phys.\  {\bf 114} (2005) 1083 
  [arXiv:hep-th/0507073].
}
\lref\TTU{
  T.~Takayanagi, S.~Terashima and T.~Uesugi,
  ``Brane-antibrane action from boundary string field theory,''
  JHEP {\bf 0103} (2001) 019 
  [arXiv:hep-th/0012210].
}
\lref\VW{
  M.~Van Raamsdonk and K.~Whyte,
  ``A light scalar particle from strong dynamics in a new holographic model of QCD,''
  arXiv:0912.0752 [hep-th].
}
\lref\Wi{
  E.~Witten,
  ``Anti-de Sitter space, thermal phase transition, and confinement in  gauge theories,''
  Adv.\ Theor.\ Math.\ Phys.\  {\bf 2} (1998) 505 
  [arXiv:hep-th/9803131].
}
\lref\WH{
  X.~J.~Wang and S.~Hu,
  ``Intersecting branes and adding flavors to the Maldacena-Nunez background,''
  JHEP {\bf 0309} (2003) 017
  [arXiv:hep-th/0307218].
}


\Title{\vbox{\baselineskip12pt 
    \hbox{IHES/P/10/09}
}} 
{\vbox{\centerline{Intersecting D4-branes Model of Holographic QCD}
\vskip12pt\centerline{and Tachyon Condensation}}} 
 
\centerline{Shigenori Seki\footnote{$^*$}{\tt sigenori@ihes.fr}} 
\bigskip 
\centerline{\it Institut des Hautes {\'E}tudes Scientifiques} 
\centerline{\it Le Bois-Marie 35, Route de Chartres, Bures-sur-Yvette 91440, France} 
 
 
\vskip .3in 
 
\centerline{\bf Abstract} 

We consider the intersecting D4-brane and anti-D4-brane model 
of holographic QCD, 
motivated by the model that has recently been suggested 
by Van Raamsdonk and Whyte. 
We analyze such D4-branes by the use of the action 
with a bi-fundamental ``tachyon'' field, 
so that we find the classical solutions describing 
the intersecting D4-branes and the U-shaped D4-branes. 
We show that the ``tachyon'' field in the bulk theory 
provides a current quark mass and a quark condensate 
to the dual gauge theory 
and that the lowest modes of mesons obtain mass via tachyon condensation. 
Then evaluating the properties of a pion, 
one can reproduce Gell-Mann-Oakes-Renner relation. 

\Date{15 March 2010}

\newsec{Introduction}

Holography is the powerful tool for analyzing physics of strong coupling. 
Since AdS/CFT correspondence, which is a realization of holography, 
was suggested by \Ma, 
many physicists have been trying to understand QCD from the 
viewpoint of a higher dimensional bulk theory. 
For this purpose, AdS/QCD model has been proposed 
as bottom up approach \EKSS, while many models have been constructed by D-brane configurations in string theory or non-critical string theory as top down approach. 
However these models are the holographic models of the theories 
close to QCD, and we have not known the holographic model of real QCD yet. 

One of the most successful models for holographic QCD is Sakai-Sugimoto (SS) model \refs{\SS,\SSi}. 
It consists of the $N_c$ color D8-branes and 
the $N_f$ flavor D8-branes and anti-D8-branes 
whose configuration is shown in the following table: 
\bigskip 
\centerline{\vbox{\offinterlineskip 
\halign{ 
\hskip2pt \hfil#\strut\hfil \hskip2pt 
& \hfil\vrule# \hskip1pc
& \hfil#\strut\hfil \hskip1pc
& \hfil#\strut\hfil \hskip1pc 
& \hfil#\strut\hfil \hskip1pc 
& \hfil#\strut\hfil \hskip1pc 
& \hfil#\strut\hfil \hskip1pc 
& \hfil#\strut\hfil \hskip1pc
& \hfil#\strut\hfil \hskip1pc 
& \hfil#\strut\hfil \hskip1pc 
& \hfil#\strut\hfil \hskip1pc 
& \hfil#\strut\hfil \hskip2pt \cr 
\noalign{\hrule} 
&& 0 & 1 & 2 & 3 & 4 & 5 & 6 & 7 & 8 & 9 \cr 
\noalign{\hrule \vskip1pt \hrule}
$N_c$ D4 && $\circ$ & $\circ$ & $\circ$ & $\circ$ & $\circ$ & & & & \cr 
$N_f$ D8/$\overline{\rm D8}$ && $\circ$ & $\circ$ & $\circ$ & $\circ$ & & $\circ$ & $\circ$ & $\circ$ & $\circ$ & $\circ$ \cr 
\noalign{\hrule} 
} 
}}\nobreak 
\centerline{Table 1: The D-brane configuration of Sakai-Sugimoto model.} 
\bigskip\noindent 
The color D4-branes induce the $U(N_c)$ dual gauge theory, which has been studied by \Wi.
In large $N_c$, the color D4-branes are regarded as the background, 
in which the D8-branes and the anti-D8-branes are connected with each other 
and are transformed to the $N_f$ U-shaped D8-branes. 
This implies the chiral symmetry breaking from $U(N_f) \times U(N_f)$ to $U(N_f)$, 
which gives rise to a pion as a massless goldstone boson. 
In SS model the U-shaped D8-branes have been studied 
by the use of a Dirac-Born-Infeld (DBI) action, 
and a lot of properties of mesons and baryons have been calculated. 
As a result, they are in good agreement with experiments. 
On the other hand, one of the serious differences 
between SS model and the real world is 
that this model is the holographic dual of the massless QCD. 
In order to improve this problem, several works have been done: 
for instance, ones incorporated a bi-fundamendal field in terms of a tachyonic DBI action \refs{\BSS,\DN}, another added an open Wilson line \AK\ and so on \refs{\HHM,\HHLY}. 
These modifications have succeeded in introducing a current quark mass, 
so that the pion mass is recovered. 

Without the $U(N_f) \times U(N_f)$ bi-fundamental ``tachyon'' field, 
the tachyonic DBI action of the D8-brane and anti-D8-brane 
considered in \BSS\ is reduced to the D8-brane action 
of SS model in the limit of non-compact background. 
Since the bi-fundamental field has the same representation 
as the quark bi-linear $q {\bar q}$, 
the current quark mass and the quark condensate in the boundary theory 
correspond to the bi-fundamental field in the bulk theory. 
The U-shaped D8-brane has been found as the classical solution of the tachyonic DBI action. 
This solution implies that the D8-brane and the anti-D8-brane 
are annihilated with each other around the origin 
because of tachyon condensation, 
so that they combine into the single U-shaped D8-brane. 
The world-volume gauge fields on the D8-brane and the anti-D8-brane
provide us with vector mesons, axial-vector mesons and pseudo-scalar mesons. 
The pion, the lowest mode of the pseudo-scalar, becomes massive on account of the tachyon field. 
Furthermore, as a byproduct, it has been shown that the pion satisfies 
Gell-Mann-Oakes-Renner (GOR) relation \GOR\ up to a numerical factor. 

The models with flavor D4-branes, instead of D8-branes, have also been 
studied in six-dimensional non-critical string theory \refs{\CPS,\GP}. 
Furthermore recently the D4-branes model in critical string theory has been 
suggested by Van Raamsdonk and Whyte (VW) \VW. 
It consists of $N_c$ color D4-branes and $N_f$ flavor D4-branes (see Table 2). 
\bigskip 
\centerline{\vbox{\offinterlineskip 
\halign{ 
\hskip2pt \hfil#\strut\hfil \hskip2pt 
& \hfil\vrule# \hskip1pc
& \hfil#\strut\hfil \hskip1pc
& \hfil#\strut\hfil \hskip1pc 
& \hfil#\strut\hfil \hskip1pc 
& \hfil#\strut\hfil \hskip1pc 
& \hfil#\strut\hfil \hskip1pc 
& \hfil#\strut\hfil \hskip1pc
& \hfil#\strut\hfil \hskip1pc 
& \hfil#\strut\hfil \hskip1pc 
& \hfil#\strut\hfil \hskip1pc 
& \hfil#\strut\hfil \hskip2pt \cr 
\noalign{\hrule} 
&& 0 & 1 & 2 & 3 & 4 & 5 & 6 & 7 & 8 & 9 \cr 
\noalign{\hrule \vskip1pt \hrule}
$N_c$ D4 && $\circ$ & $\circ$ & $\circ$ & $\circ$ & $\circ$ & & & & \cr 
$N_f$ D4 && $\circ$ & $\circ$ & $\circ$ & $\circ$ & & $\circ$ & & & & \cr 
\noalign{\hrule} 
} 
}}\nobreak 
\centerline{Table 2: The D-brane configuration of Van Raamsdonk-Whyte model.} 
\bigskip\noindent 
Both of VW model and SS model have the same gauge sector 
by the color D4-branes as \Wi.
On the other hand, since the codimension 
of the flavor D4-branes is bigger than the one of the D8-branes, 
there exist more scalar (meson) fields in VW model than in SS model. 
In \VW, the mass spectra of the scalar mesons have been numerically evaluated 
and the relations between the mass spectra and the constituent quark mass
have been clarified. 
The constituent quark mass is determined 
by the IR boundary condition of the flavor D4-branes. 
Note that the constituent quark mass in SS model has been 
introduced by the non-antipodal generalization \refs{\CPS,\ASY,\PSZ}. 

In this paper we are interested in incorporating a current quark mass into VW model. 
We shall consider the tachyonic DBI action of 
the $N_f$ flavor D4-branes and anti-D4-branes. 
This action is reduced to VW model, if the tachyon field vanishes everywhere. 
The D4-branes and the anti-D4-branes which we shall consider 
are also regarded as the intersecting D4-branes, 
and there exists a tachyon around the intersection point. 
The recombination of intersecting D-branes has been studied by \refs{\HN,\JL}. 
In our case, the intersecting D4-branes should be recombined 
by the tachyon condensation and become U-shaped D4-branes 
in the way similar to \BSS. 
Then the tachyon field corresponds to the quark bi-linear 
and the pion obtains mass. 
In \refs{\WH,\PR}, the other kinds of intersecting D-brane systems have been studied 
for adding flavors into the holographic models. 

For simplicity, we shall consider the $N_f=1$ case. 
In Section 2, we shall explain the background metric used in \VW. 
Then the tachyonic action of the D4-brane and anti-D4-brane 
in the non-compact limit of this background will be analyzed, 
and we shall find classical solutions in the IR and UV asymptotic regions. 
In Section 3 we shall evaluate the quark mass and the quark condensate  from the bulk theory in terms of those solutions. 
In Section 4 we shall consider the fluctuations 
of the gauge fields on the world-volume of flavor D4-branes, 
and in Section 5 some properties of the pion which is derived from this gauge field will be calculated. 
We shall also show that those properties satisfy GOR relation. 
In Section 6, we shall give some comments on scalar fields. 
Section 7 is devoted to the conclusions and comments.

\newsec{Analyses of flavor D4-brane and anti-D4-brane}

\subsec{$N_c$ color D4-branes as background}

When $N_c \gg N_f\,(=1)$, the $N_c$ color D4-branes are regarded 
as the background, 
in which we can treat the flavor D4-brane and anti-D4-brane as probes. 
The metric of this background is described as
\eqnn\Dfourmet
$$\eqalignno{
&ds^2 = \biggl({U \over R}\biggr)^{3 \over 2}
	\bigl[ \eta_{\mu\nu} dx^\mu dx^\nu
	+f(U)dx_4^2 \bigr] 
	+\biggl({R \over U}\biggr)^{3 \over 2}
		\biggl[{dU^2 \over f(U)} +U^2 d\Omega_4^2\biggr] \,, \cr
&\quad	f(U) := 1 - \biggl({U_{\rm KK} \over U}\biggr)^3 \,,\quad
	e^\phi = g_s \biggl({U \over R}\biggr)^{3 \over 4} \,,\quad
	F_4 = {2\pi N_c \over V_4} \epsilon_4 \,, &\Dfourmet
}$$
where the volume of a unit four-sphere $V_4$ is equal to $8\pi^2/3$. 
Since the $x_4$ direction is compact and $U \geq U_{\rm KK}$, 
this background has a cigar geometry. 
The period of $x_4$ is determined by the smoothness at $U=U_{\rm KK}$ 
and leads to the Kaluza-Klein mass 
$$
M_{\rm KK} = {3 \over 2}\sqrt{U_{\rm KK} \over R^3} \,.
$$
$R$ and the 't Hooft coupling $\lambda\ (= g_{\rm YM}^2 N_c)$ 
are denoted in terms of 
the string coupling $g_s$ and the string scale $l_s$ by 
$$
R^3 = \pi N_c g_s l_s^3 \,, \quad 
\lambda = 2\pi M_{\rm KK} N_c g_s l_s \,.
$$
We shall use the coordinate transformation introduced by \VW, 
\eqn\ccdiff{
{d\rho \over \rho} = {dU \over \sqrt{f(U)} U} \,. 
}
This differential equation can be solved 
\eqn\coorchange{
2\biggl({U \over U_{\rm KK}}\biggr)^{3 \over 2} 
= \biggl({\rho \over \rho_{\rm KK}}\biggr)^{3 \over 2} 
	+\biggl({\rho_{\rm KK} \over \rho}\biggr)^{3 \over 2} \,.
}
Since \ccdiff\ is invariant under the rescaling of $\rho$, 
we fix the scale of $\rho$ so that 
$$
\rho_{\rm KK} = 2^{-{2 \over 3}}U_{\rm KK} \,,
$$
for the later convenience. 
Since the coordinate $U$ in \Dfourmet\ has the lower bound $U_{\rm KK}$ 
and the right hand side of \coorchange\ is the monotonically increasing function of $\rho$, 
$\rho_{\rm KK}$ is the lower bound of the coordinate $\rho$. 
Then the metric \Dfourmet\ is rewritten as 
\eqnn\rhomet
$$\eqalignno{
&ds^2 = \biggl({\rho \over R}\biggr)^{3 \over 2} g_+(\rho) \biggl[
	\eta_{\mu\nu}dx^\mu dx^\nu 
	+\biggl({g_-(\rho) \over g_+(\rho)}\biggr)^2 dx_4^2 \biggr] 
	+\biggl({R \over \rho}\biggr)^{3 \over 2}g_+(\rho)^{1 \over 3}
	\bigl[ d\rho^2  + \rho^2 d\Omega_4^2 \bigr] \,, \cr
&\quad	g_\pm (\rho) := 1 \pm \biggl({\rho_{\rm KK} \over \rho}\biggr)^3 \,, \quad 
	e^\phi = g_s \biggl({\rho \over R}\biggr)^{3 \over 4}\sqrt{g_+(\rho)} \,. &\rhomet 
}$$
Note that $d\rho^2 +\rho^2 d\Omega_4^2$ is the metric of $\Bbb{R}^5$. 
The $x_4$ direction shrinks at $\rho = \rho_{\rm KK}$ because of $g_-(\rho_{\rm KK}) = 0$. 

Here we shall consider the non-compact limit, that is, $U_{\rm KK} \to 0$. 
Since $\rho_{\rm KK}$ also goes to zero in this limit, 
the metric and dilaton \rhomet\ are reduced to 
\eqn\ncmet{
ds^2 = \biggl({\rho \over R}\biggr)^{3 \over 2} \bigl[
	\eta_{\mu\nu}dx^\mu dx^\nu 
	+ dx_4^2 \bigr] 
	+\biggl({R \over \rho}\biggr)^{3 \over 2} 
	\bigl[ d\rho^2  + \rho^2 d\Omega_4^2 \bigr] \,,\quad 
e^\phi = g_s \biggl({\rho \over R}\biggr)^{3 \over 4} \,. 
}

\subsec{D$p$-brane-anti-D$p$-brane action}

The action of D$p$-brane and anti-D$p$-brane consists of the DBI action and 
the Chern-Simons action, 
\eqn\dadaction{
S = S_{\rm DBI} + S_{\rm CS} \,.
}

The tree-level effective DBI action $S_{\rm DBI}$
has been given by \Ga. 
Without NSNS B-fields, the action is written down as
\eqn\dadDBI{
S_{\rm DBI} = -T_p \int d^{p+1}\sigma \sum_{i=1}^2 
V(|\tau|) \sqrt{\CQ^{(n)}} e^{-\phi(X^{(n)})} \sqrt{-\det {\bf A}^{(n)}} \,,
}
where 
$$\eqalign{
{\bf A}^{(n)}_{ab} &= P^{(n)}\biggl[ g_{ab} - {|\tau|^2 \over 2\pi\alpha' \CQ^{(n)}} g_{ai}l^i l^j g_{jb} \biggr] 
	+2\pi\alpha' F_{ab}^{(n)} \cr
&\quad +{1 \over \CQ^{(n)}}\biggl( 
	\pi\alpha' (D_a\tau (D_b \tau)^* +D_b\tau (D_a\tau)^*) \cr
&\qquad	+{i \over 2}(g_{ai} + \partial_a X^{(n)j}g_{ji})l^i(\tau(D_b\tau)^* -\tau^*D_b\tau) \cr
&\qquad	+{i \over 2}(\tau(D_a\tau)^* -\tau^*D_a\tau) l^i (g_{ib} - g_{ij}\partial_b X^{(n)j}) \biggr) \,, \cr
\CQ^{(n)} &= 1 + {|\tau|^2 \over 2\pi\alpha'}l^i l^j g_{ij}(X^{(n)}) \,.
}$$
$T_p$ is the tension of D$p$-brane. 
$a,b$ denote the tangent directions of D-branes, 
while $i,j$ denote the transverse ones. 
The D$p$-brane and the anti-D$p$-brane are labelled by 
$n=1$ and 2 respectively. 
Then the separation between these D-branes is defined by $l^i = X^{(1)i} - X^{(2)i}$.
$P^{(n)}$ means the pullback from the target space to the world-volume, 
that is, $P^{(n)}[\eta_{ab}] = \eta_{MN}\partial_a X^{(n)M} \partial_b X^{(n)N}$, where $M,N$ are $0,\dots, 9$.  
$\tau$ is a bi-fundamental ``tachyon'' field, 
of which covariant derivative is denoted by 
$D_a \tau = \partial_a \tau -i (A^{(1)}_a - A^{(2)}_a) \tau$.
We take the gauge $\Im \tau = 0$ 
and define the ``tachyon'' field $T$ by $\Re \tau = T$. 
The tachyon potential is given by \refs{\BLW\BW\LP{--}\LLM}
\eqn\tacpot{
V(T) = {1 \over \cosh (\sqrt{\pi}T)} \,.
}
Though the other candidate of the tachyon potential is $V(T) = e^{-T^2/4}$ \KMM, we adopt \tacpot\ in this paper by following \Ga.

When there are RR fields in a background, we have to consider the Chern-Simons action. 
Our background \Dfourmet\ has two non-vanishing RR fields, 
which are the three-form $C_3$ ($F_4 = dC_3$) and the five-form $C_5$. 
In this paper we shall use the same notation as \refs{\SS,\Po} (see Appendix A in \SS). 
We calculate $dC_5$ from \Dfourmet, 
$$
dC_5 = {1 \over (2\pi)^2\alpha'}*\! dC_3 
= {N_c \over 2\pi\alpha' V_4}{U^2 \over R^6} dx^0\wedge dx^1 \wedge dx^2 \wedge dx^3 \wedge dx^4 \wedge dU \,.
$$
Then $C_5$ is evaluated as 
\eqn\RRfiveform{
C_5 = {N_c \over 2\pi\alpha' V_4}{U^3 - U_{\rm KK}^3 \over 3R^6} dx^0\wedge dx^1 \wedge dx^2 \wedge dx^3 \wedge dx^4 
}
where the integration constant is determined 
so that the singularity at $U=U_{\rm KK}$ is removed \VW. 
In the non-compact limit \ncmet, the five-form becomes 
$$
C_5 = {N_c \over 16\pi^3\alpha' R^6}\rho^3 dx^0\wedge dx^1 \wedge dx^2 \wedge dx^3 \wedge dx^4 \,,
$$
where we used $V_4 = 8\pi^2/3$. 
The RR fields appear in the Chern-Simons action as the pullback to the world-volume, $P[C_q]$. 
The world-volume coordinates of the flavor D4-brane and anti-D4-brane 
that we shall consider are identified with 
the dual gauge theory directions $x^\mu$ ($\mu=0,1,2,3$) 
and one of the directions of $\rho$ and $\Omega_4$, 
and the collective coordinates of the flavor branes 
are independent of $x^\mu$. 
Therefore the pullback of $C_3$ vanishes, and 
we can concentrate on only the five-form $C_5$. 

For the coincident D$p$-brane and anti-D$p$-brane the Chern-Simons action 
including the coupling with the bi-fundamental tachyon field has been 
proposed by \KW. 
It is described in terms of a curvature supermatrix $\CF$ as 
$$
S_{\rm CS} = \int_{\Sigma_{p+1}} P[C]\wedge {\rm Str}\exp(i2\pi\alpha'\CF) \,,
$$
where 
$$\eqalign{
&C=\sum_n (i2\pi\alpha')^{-{p-n+1 \over 2}}(2\pi)^{-{p-n+1 \over 2}}C_n \,,\cr
&i2\pi\alpha'\CF = \pmatrix{i2\pi\alpha'F^{(1)} -T^*T & DT^* \cr
				DT & i2\pi\alpha'F^{(2)} -TT^*} \,.
}$$
This action has been proved by \refs{\KL,\TTU} from the viewpoint of the boundary string field theory.\foot{Other progress on the Chern-Simons actions 
for the D-brane-anti-D-brane or non-BPS D-branes are made by \GH.} 
If the flavor D4-brane and anti-D4-brane coincide 
with each other in our background, 
the RR field does not couple with the world-volume gauge field and the tachyon field. 
Because the only relevant RR field is the five-form $C_5$. 
By the same reason, although the Chern-Simons action 
for the separated D$p$-brane and anti-D$p$-brane 
with the tachyon has not been well established, 
we can easily infer that the Chern-Simon action for the separated D4-brane 
and anti-D4-brane in our background is  
\eqn\dadCS{
S_{\rm CS} = \int_{\rm D4} P^{(1)}[C_5] - \int_{\overline{\rm D4}} P^{(2)}[C_5] \,.
}
The minus sign is due to the supertrace\foot{${\rm Str}\pmatrix{A&B \cr C&D} \equiv \tr A - \tr D$.} 
and implies the relative orientation of the D4-brane and anti-D4-brane. 

\subsec{Intersecting D4-branes}

Let us change the spherical coordinates of the $\Bbb{R}^5$ part of \ncmet\ 
to the cylindric ones, 
$$
d\rho^2 + \rho^2 d\Omega_4^2 = dr^2 +r^2d\theta^2 +d{\vec x}_T^2 \,, \quad 
\rho^2 = r^2 + |{\vec x}_T|^2 \,.
$$
${\vec x}_T$ denotes the three-dimensional space transverse 
to the $r$-$\theta$ plane. 
$r$ is equal to $\rho$ on the plane defined by ${\vec x}_T = 0$, 
where the D4-brane and anti-D4-brane that we shall consider 
are located.\foot{The $\Bbb{R}^5$ part of the metric can be written also by the Cartesian coordinates as $dy^2 + dz^2 +d{\vec x}_T^2$, 
where $\rho^2 = y^2 + z^2 + {\vec x}_T^2$. 
Though it seems possible to put the parallel D4-brane and anti-D4-brane separated along the $z$ direction by the analogy of \BSS, this parallel configuration is not a classical solution. Details are explained in Appendix A.} 

The world-volume coordinates of the D4-brane and anti-D4-brane 
are denoted by $(x^0,x^1,x^2,x^3,r)$, 
while we set the ansatz of the embeddings of these D-branes 
to be $x_4^{\rm D4}=x_4^{\rm \overline{D4}} = 0$, $\theta^{\rm D4} = -\theta^{\rm \overline{D4}} = \Theta(r)/2$ 
and ${\vec x}_T^{\rm D4}= {\vec x}_T^{\rm \overline{D4}} = 0$. 
We shall consider the only $r$ dependence of the tachyon field $T$ for simplicity. 
Substituting these ansatz into \dadDBI,  
we obtain the tachyonic DBI action, 
\eqna\itsaction
$$\eqalignno{
S_{\rm DBI} &= -{2T_4 \over g_s}\int d^4x dr\, V(T) 
	\biggl({r \over R}\biggr)^{3 \over 2} 
	\sqrt{\CD} \,, &\itsaction{a} \cr
&\CD = 1+{r^2 \over 4}\Theta'^2 
	+2\pi\alpha'\biggl({r \over R}\biggr)^{3 \over 2}T'^2
	+{r^2 \over 2\pi\alpha'}\biggl({R \over r}\biggr)^{3 \over 2} \Theta^2 T^2 \,,  &\itsaction{b} 
}$$
where $'$ denotes the derivative with respect to $r$. 
We should note that the Chern-Simon action \dadCS\ vanishes 
due to the ansatz $x_4^{\rm D4}=x_4^{\rm \overline{D4}} = 0$. 
We shall give in Section 7 further discussion 
about the more general ansatz of $x^4$ direction, 
in which the Chern-Simons action contributes.  
Finally in our setup the action of the D4-brane and anti-D4-brane 
consists of only the DBI action \itsaction{}, 
and the equations of motion for $\Theta(r)$ and $T(r)$ are written down as  
\eqna\itseom
$$\eqalignno{
&{d \over dr}\biggl[V(T) \biggl({r \over R}\biggr)^{3 \over 2} {r^2 \over 4\sqrt{\CD}} \Theta' \biggr] = {V(T) r^2 \over 2\pi\alpha'\sqrt{\CD}} \Theta T^2 \,, &\itseom{a} \cr
&{d \over dr}\biggl[V(T) \biggl({r \over R}\biggr)^3 {2\pi\alpha' \over \sqrt{\CD}} T' \biggr] 
= {V(T) r^2 \over 2\pi\alpha' \sqrt{\CD}} \Theta^2 T +\biggl({r \over R}\biggr)^{3 \over 2}\sqrt{\CD}{dV(T) \over dT} \,. &\itseom{b}
}$$
We can easily find the trivial solution of these equations, 
\eqn\trivsol{
\Theta(r) = \Theta_\infty \,, \quad T(r) = 0 \,,
}
where $\Theta_\infty$ is the constant determined by a boundary condition. 
If we regard $(r,\theta)$ as the Cartesian coordinates, 
this solution describes the parallel D4-brane and anti-D4-brane, 
which are the analogues of the parallel D8-brane and anti-D8-brane in \BSS. 
\trivsol\ is also realized as the D4-branes intersecting at $r=0$ 
in the polar coordinates (fig.~1(a)). 
Simultaneously the tachyon field stays at the top of the potential $V(T)$ for any $r$.
\bigskip 
\vbox{\centerline{\epsfbox{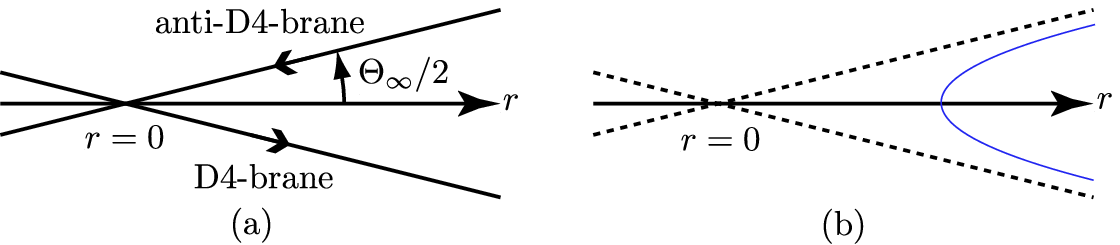}} 
\centerline{\fig\figintdfour{} (a) The intersecting D4-branes. (b) The U-shaped D4-brane.}} 
\bigskip\nobreak\noindent 

We shall search a non-trivial solution. 
The bi-fundamental ``tachyon'' field $T(r)$ is really tachyonic around $r=0$, 
in which the D4-brane and the anti-D4-brane would be annihilated 
with each other via the tachyon condensation. 
If $\Theta_\infty$ is sufficiently small, 
then the flavor D4-brane and anti-D4-brane prefer to recombine 
with each other rather than with the color D4-branes. 
We can expect that as a result 
the flavor D4-branes become the U-shaped D4-branes \refs{\HN,\JL}  
and the tip of the U-shape is far apart from the origin $r=0$ 
where the color D4-branes are originally located (see fig.~1(b)). 
Since at present it is difficult to fully solve the equations of motion \itseom{}
analytically, we shall concentrate on the IR and UV asymptotic behavior. 

For later convenience, we introduce a dimensionless variable $s$ 
by rescaling $r$ as 
$$
r = {(2\pi\alpha')^2 \over R^3} s \,. 
$$
The action \itsaction{} and the equations of motion \itseom{} 
are rewritten as 
\eqna\dimlessact
\eqna\dimlesseom
$$\eqalignno{
S &= -{2(2\pi\alpha')^5 T_4 \over g_s R^9}\int d^4x ds\, 
	V(T) s^{3 \over 2} \sqrt{\CD} \,, &\dimlessact{a} \cr
&\CD = 1 + {1 \over 4}s^2 \biggl({d\Theta \over ds}\biggr)^2 + s^{3 \over 2} \biggl({dT \over ds}\biggr)^2 
	+s^{1 \over 2} \Theta^2 T^2 \,, &\dimlessact{b} 
}$$
and
$$\eqalignno{
&{1 \over 4}{d \over ds} \biggl[
{V(T) \over \sqrt{\CD}} s^{7 \over 2} {d\Theta \over ds} \biggr]
= {V(T) \over \sqrt{\CD}} s^2 \Theta T^2 \,, &\dimlesseom{a} \cr
&{d \over ds} \biggl[
{V(T) \over \sqrt{\CD}} s^3 {dT \over ds} \biggr] 
= {V(T) \over \sqrt{\CD}} s^2 \Theta^2 T
	+ {dV(T) \over dT} s^{3 \over 2} \sqrt{\CD} \,. &\dimlesseom{b} 
}$$
These equations are useful for numerical analyses which will be done 
in the following subsections.

\subsec{Solutions in the IR region}

Since the D4-brane and the anti-D4-brane are sufficiently close 
to each other in the IR region $(s \ll 1)$, 
the open string stretched between these D-branes has a tachyonic mode, 
that is to say, the ``tachyon'' field $T$ becomes really a tachyon. 
We expect that the condensation of this tachyon 
gives rise to the recombination of the D-branes, 
so that the D4-brane and anti-D4-brane become U-shaped D4-branes. 
We shall look for such a solution in terms of the power expansion 
of $\Theta(s)$ and $T(s)$ around $s = s_0 (\ll 1)$, 
\eqna\IRansatz
$$\eqalignno{
\Theta(s) &= \theta_0 (s - s_0)^a + \theta_1 (s - s_0)^{a+1} +\cdots \,, &\IRansatz{a} \cr
T(s) &= t_0 (s - s_0)^b + t_1 (s - s_0)^{b+1} +\cdots \,. &\IRansatz{b}
}$$
$s_0$ is the constant determined by a boundary condition. 
Substituting these ansatz into the equations of motion \dimlesseom{}, 
we find 
\eqn\solpower{
0 < a < 1  \,, \quad 
b = -2 \,, \quad 
t_0={\sqrt{\pi}  \over 2}s_0^{3 \over 2}a \,.
}
$a < 1$ is required for the smoothness of the D4-brane 
at $s = s_0$. 
We also write down this IR asymptotic solution 
in the original variable $r$: 
\eqna\IRsolution
$$\eqalignno{
\Theta(r) &= {\theta_0 R^{3a} \over (2\pi\alpha')^{2a}} (r-r_0)^a +\CO\bigl((r-r_0)^{a+1}\bigr) \,, &\IRsolution{a} \cr
T(r) &= {(2\pi\alpha')\sqrt{\pi} a r^{3/2} \over 2 R^{3/2}} (r-r_0)^{-2} +\CO\bigl((r-r_0)^{-1}\bigr) \,, &\IRsolution{b}
}$$
where $r_0 = (2\pi\alpha')^2 s_0 /R^3$. 

We shall analyze the IR behavior 
by solving numerically the equations of motion \dimlesseom{}. 
Referring to the solution \solpower, we set by hand the IR boundary 
$s_0 = 10^{-4}$ and the following IR initial conditions:
$$
\Theta(10^{-4}) = 10^{-5} \,, \quad {d\Theta \over ds}(10^{-4}) = 10^6 \,,\quad 
T(10^{-4}) = 10^3 \,, \quad {dT \over ds}(10^{-4}) = -10^6 \,.
$$
Though \solpower\ implies that $\Theta(s_0) = 0$ 
and ${d\Theta \over ds}(s_0) = T(s_0) = -{dT \over ds}(s_0) = \infty$, 
the numerical computation does not proceed from such singular values. 
We therefore replaced zero and infinity with sufficiently small and large numbers respectively. 
We should also note that the ansatz \IRansatz{} is valid in the range of $(s-s_0) \ll s_0$. 
The numerical result in this range is depicted in fig.~2. 
\bigskip 
\vbox{\centerline{\epsfbox{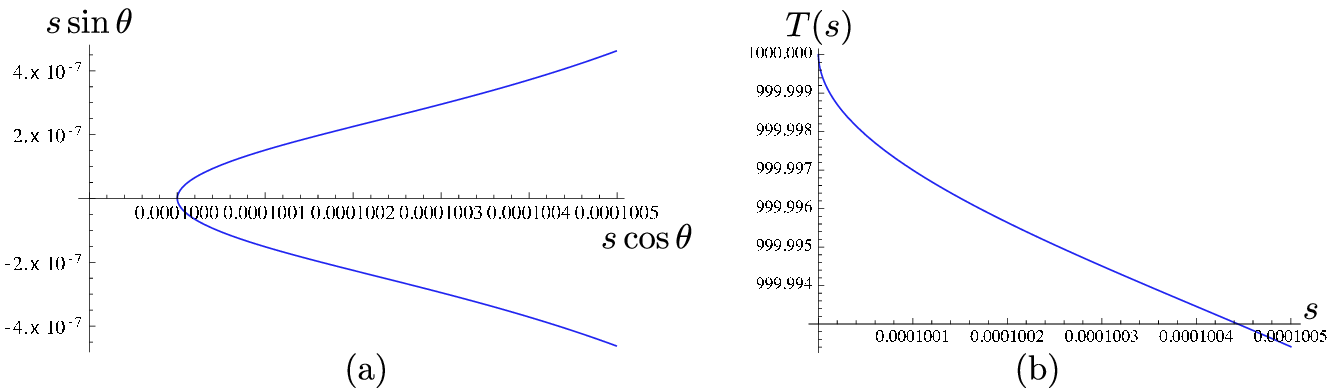}} 
\centerline{\fig\figirlimit{} (a) The polar plot of D4-brane in IR. (b) The behavior of the tachyon field in IR.}} 
\bigskip\nobreak\noindent 
The D4-branes become U-shape and the tachyon field monotonically decreases 
as we expected from the solution \solpower. 
However, in the region of larger $s$, the numerical result shown in fig.~3 
does not agree with our expectation. 
\bigskip\vbox{
\centerline{\epsfbox{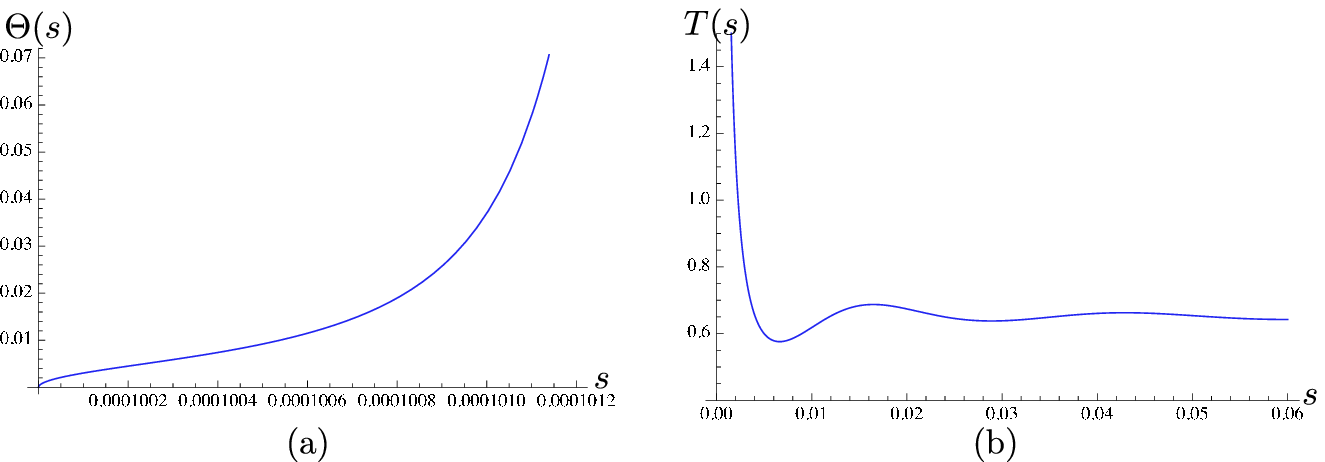}}
\centerline{\vbox{\offinterlineskip 
\halign{ \strut# & #\hfil \cr
\fig\figirLarge{} & (a) The plot of the angle $\Theta(s)$ from IR to larger $s$. \cr
 & (b) The plot of the tachyon field $T(s)$ from IR to larger $s$. \cr
}}}
}
\bigskip\nobreak\noindent 
In \figirLarge, $\Theta$ blows up 
and $T$ starts to oscillate, when $s$ becomes  large. 
The reason why such unfavorable behavior appears is that the numerical analysis is sensitive to the IR initial conditions. 
Some literatures on the numerical analyses of tachyonic DBI action 
have encountered a similar issue \refs{\BSS,\DN}. 

\subsec{Solutions in the UV region}

In this subsection we shall concentrate on the UV region $(s \gg 1)$. 
One can suppose that the D4-branes are scarcely affected 
by the tachyon condensation because the ``tachyon'' field is massive in this region. 
So we shall consider the small fluctuations from the trivial solution \trivsol, 
\eqn\UVansatz{
\Theta(s) = \Theta_\infty + \delta\Theta(s) \,,\quad 
T(s) = 0 +\delta T(s) \,.
}
Calculating the effective action from \itsaction{} in the quadratic order of these fluctuation fields, we obtain the reduced equations of motion, 
\eqna\fluceom
$$\eqalignno{
&{d \over ds}\biggl( s^{3 \over 2}{d\delta\Theta \over ds} \biggr) = 0 \,, &\fluceom{a} \cr
&{d \over ds}\biggl( s^3 {d\delta T \over ds} \biggr) = (s^{1 \over 2} \Theta_\infty^2 - \pi)s^{3 \over 2}\delta T \,. &\fluceom{b}
}$$
\fluceom{a} with the boundary condition $\delta\Theta(\infty) = 0$ 
is easily solved, 
\eqn\UVdeltasol{
\delta\Theta(s) = C_\theta s^{-{1 \over 2}} \,,
}
where $C_\theta$ is an integration constant. 
On the other hand, it is hard to solve \fluceom{b} analytically. 
Since we can approximate the equation of motion \fluceom{b} by  
${d \over ds}\bigl( s^3 {d\delta T \over ds}\bigr) = \Theta_\infty^2 s^2 \delta T$ in the large $s$ region that we are now considering, 
the classical solution of the tachyon field is described as   
\eqn\UVtacsol{
\delta T(s) = {C_{\rm nn} \over \Theta_\infty^2}{I_2(2\Theta_\infty\sqrt{s}) \over s}
	+{C_{\rm n} \over \Theta_\infty^2}{K_2(2\Theta_\infty\sqrt{s}) \over s} \,.
}
$I_n(z)$ and $K_n(z)$ are the modified  Bessel functions of the 
first and the second kind respectively. 
$C_{\rm nn}$ and $C_{\rm n}$ are integration constants. 
Since the first term in \UVtacsol\ diverges at the limit 
$s \to \infty$, it implies a non-normalizable mode. 
On the other hand, the second term converges 
and corresponds to a  normalizable mode. 
Since we used the perturbation around $T=0$, 
the solution \UVtacsol\ is valid under
$C_{\rm nn}I_2(2\Theta_\infty\sqrt{s_\infty}) \lesssim C_{\rm n}K_2(2\Theta_\infty\sqrt{s_\infty})$, where $s_\infty$ is a cutoff parameter.  
For the later use, here we write down the UV asymptotic solution in terms of $r$, 
\eqna\UVsolution
$$\eqalignno{
\Theta(r) &= \Theta_\infty + {2\pi\alpha' C_\theta \over R^{3/2}} {1 \over \sqrt{r}}\,, &\UVsolution{a} \cr
T(r) &= {(2\pi\alpha')^2 \over R^3 \Theta_\infty^2}{1 \over r} \biggl[
	C_{\rm nn} I_2\biggl({\Theta_\infty R^{3/2} \over \pi\alpha'}\sqrt{r}\biggr)
	+C_{\rm n} K_2\biggl({\Theta_\infty R^{3/2} \over \pi\alpha'}\sqrt{r}\biggr) \biggr] \,. &\UVsolution{b}
}$$

We shall analyze the UV behavior numerically.  
Following the solutions \UVdeltasol\ and \UVtacsol, 
we impose the initial conditions:
$$
\Theta(10^4) = 0.5 \,, \quad 
{d\Theta \over ds}(10^4) = 10^{-10} \,, \quad 
T(10^4) = 10^{-16} \,, \quad 
{dT \over ds}(10^4) = -10^{-22} \,.
$$
Then the solutions of the equations of motion \dimlesseom{} are 
drawn in fig.~4. 
\bigskip\vbox{
\centerline{\epsfbox{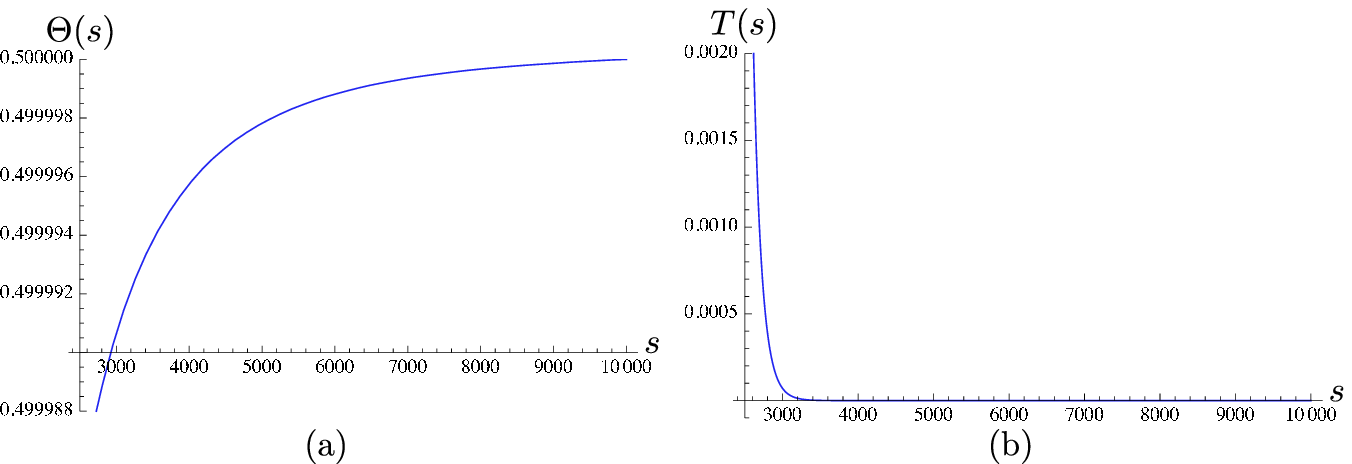}}
\centerline{\vbox{\offinterlineskip 
\halign{ \strut# & #\hfil \cr
\fig\figuvlimit{} & (a) The plot of the angle $\Theta(s)$ from UV. \cr
 & (b) The plot of the tachyon field $T(s)$ from UV. \cr
}}}
}
\bigskip\nobreak\noindent 
When $s$ becomes smaller, $\Theta(s)$ blows down. 
This implies that D4-branes would be connected at a certain small $s$. 
And simultaneously the tachyon field $T(s)$ blows up, that is to say, 
the tachyon flows to the true vacuum $(T=\infty)$ in the tachyon potential. 
These results agree with our expectation discussed so far. 

\newsec{Quark mass and condensate}

Following the usual AdS/CFT dictionary, a normalizable mode 
in a bulk theory corresponds to a physical quantity 
in the dual boundary theory, 
while a non-normalizable mode corresponds to a parameter. 
The ``tachyon'' field in the bulk theory has 
a bi-fundamental representation, 
which the quark bi-linear has in the boundary theory. 
Now let us assume 
\eqn\nnmq{
C_{\rm nn} = \Lambda m_q \,.
}
$m_q$ is a current quark mass and $\Lambda$ is a fixed parameter with a length scale.  
We should recall that the quark mass term appears in QCD Lagrangian as 
$m_q q{\bar q}$. 
So we can extract the quark condensate $\langle q{\bar q}\rangle$ by 
differentiating the energy density with respect to $m_q$ 
and then putting $m_q=0$ \KMMW. 
In our model, since the energy density $\CE$ is evaluated by  
the Euclidean action $S = -\int d^4x\, \CE$, we can calculate the quark condensate, 
\eqn\nqcond{
\langle q{\bar q} \rangle = {\delta \CE \over \delta m_q}\biggr|_{m_q = 0} 
	= \Lambda {\delta \CE \over \delta C_{\rm nn}}\biggr|_{C_{\rm nn} = 0} 
= {(2\pi\alpha')^5 T_4 \over 2 g_s R^9}{\Lambda \over \Theta_\infty^4}C_{\rm n} \,.
}
As a result, the quark condensate corresponds 
to the coefficient $C_{\rm n}$ in 
the normalizable mode and it is consistent with the AdS/CFT dictionary.

\newsec{Gauge fields}

Let us incorporate the $U(1)\times U(1)$ gauge fields 
$A_i^{(n)}(x^\mu,r)$ $(i=0,1,2,3,r)$ on the probe D4-branes into the action \itsaction{}. 
Here we redefine the gauge fields by 
$$
A^{(\pm)}_i = {1 \over 2}(A^{(1)}_i \pm A^{(2)}_i) \,.
$$
In the following subsections, we shall show 
that $A^{(+)}$ and $A^{(-)}$ correspond 
to the vector mesons 
and the axial-vector and pseudo-scalar mesons respectively. 
These gauge fields are dealt with as the perturbation on the U-shaped D4-branes which are given by the classical solutions $\Theta(r)$ and $T(r)$  discussed in Section 2. 
Note that the Chern-Simons action \dadCS\ contains no gauge field. 
We calculate the effective action of $A^{(\pm)}$ at quadratic order, 
\eqnn\gaugeaction
$$\eqalignno{
S_{\rm gauge}[A^{(\pm)}] &= -\int d^4x dr \Bigl[
	\CC_1\Bigl( \bigl|F^{(+)}_{\mu\nu}\bigr|^2 + \bigl|F^{(-)}_{\mu\nu}\bigr|^2 \Bigr)
	+\CC_2\Bigl( \bigl|F^{(+)}_{\mu r}\bigr|^2 + \bigl|F^{(-)}_{\mu r}\bigr|^2 \Bigr) \cr
&\quad	+\CC_3 \bigl|A^{(-)}_\mu\bigr|^2
	+\CC_4 \bigl|A^{(-)}_r \bigr|^2
	+\CC_5 F^{(-)}_{\mu r}A^{(-)\mu} 
\Bigr] \,. &\gaugeaction 
}$$
It is remarkable that the mass term of $A^{(-)}$ appears explicitly in this action. 
The coefficients $\CC_a$ ($a=1,\dots, 5$) are the functions of $r$ given by 
\eqnn\gactcoeff
$$\eqalignno{
&\CC_1 = {(2\pi\alpha')^2 T_4 \over 2g_s} V(T) \biggl({R \over r}\biggr)^{3 \over 2} \sqrt{\CD} \,, \quad
\CC_2 = {(2\pi\alpha')^2 T_4 \over g_s} V(T) \biggl({r \over R}\biggr)^{3 \over 2} {\CQ \over \sqrt{\CD}} \,, \cr
&\CC_3 = {8\pi\alpha' T_4 \over g_s}V(T) {\sqrt{\CD} \over \CQ}T^2
	+{T_4 \over g_s}V(T){r^4 \over \CQ \sqrt{\CD}}\biggl({R \over r}\biggr)^{3 \over 2}\Theta^2 \Theta'^2 T^4 \,, \cr
&\CC_4 = {8\pi\alpha' T_4 \over g_s}V(T) \biggl({r \over R}\biggr)^3{1 \over \sqrt{\CD}}T^2 \,, \quad
\CC_5 = {4\pi\alpha' T_4 \over g_s}V(T){r^2 \over \sqrt{\CD}} \Theta \Theta' T^2 \,, &\gactcoeff
}$$ 
where 
$\CQ = 1 + (2\pi\alpha')^{-1} R^{3/2} r^{1/2} \Theta^2 T^2$
and $\CD$ has been defined by \itsaction{b}. 
Without the tachyon field ($T=0$), 
the coefficients $\CC_3, \CC_4, \CC_5$ vanish, 
that is to say, all the gauge fields $A^{(\pm)}$ have no mass term. 

\subsec{$A^{(+)}$ part}

In terms of the $U(1)$ gauge symmetry, we can fix $A^{(+)}_r = 0$. 
Under this gauge, the $A^{(+)}$ part of 
the action \gaugeaction\ is described as 
\eqn\Aplusact{
S^{(+)} = - \int d^4x dr \Bigl[
	\CC_1 \bigl| F^{(+)}_{\mu\nu} \bigr|^2 
	+\CC_2 \bigl| A^{(+)}_\mu{}' \bigr|^2 
	\Bigr] \,.
}
We consider the mode expansion of the gauge field $A^{(+)}_\mu$, 
\eqn\Aplusmode{
A^{(+)}_\mu(x^\mu,r) = \sum_n a^{(+)}_{n\mu}(x^\mu) \psi_n(r) \,.
}
Each mode $\psi_n$ is determined by the eigen equation
\eqn\Apluseigeneq{
-{1 \over 2}{d \over dr}(\CC_2 \psi_n{}') 
	= \bigl(m^{(+)}_n\bigr)^2 \CC_1 \psi_n \,,
}
while we set the normalization condition 
\eqn\Aplusnorm{
{1 \over 4}\delta_{mn} = \int dr\, \CC_1 \psi_m\psi_n \,.
}
Then the action \Aplusact\ is reduced to the four-dimensional action
$$
S^{(+)} = -\int d^4x \sum_n \biggl[ 
	{1 \over 4}\bigl| f^{(+)n}_{\mu\nu} \bigr|^2 
	+{1 \over 2} \bigl(m^{(+)}_n\bigr)^2 \bigl| a^{(+)}_{n\mu} \bigr|^2 
\biggr] \,,
$$
where $f^{(+)}_{n\mu\nu} = \partial_\mu a^{(+)}_{n\nu} - \partial_\nu a^{(+)}_{n\mu}$. 
The four-dimensional gauge field $a^{(+)}_{n\mu}$ describes a vector meson. 

We shall check in the same way as \BSS\ whether a massless vector meson exists. 
Let us assume $m^{(+)}_0 = 0$, which also implies $m_q\, (\sim C_{\rm nn}) = 0$. We can estimate $\psi_0$ 
from the eigen equation \Apluseigeneq, so that 
\eqn\Pzeromode{
\psi_0(r) \sim \int^r d{\tilde r} {1 \over \CC_2({\tilde r})} \,.
}
In the UV region, the classical solution of $\Theta(r)$ and $T(r)$ has been shown in \UVsolution{}. 
We should note that $T(r)$ is the exponentially decreasing function of $r$ 
under that assumption. 
Substituting these $\Theta(r)$ and $T(r)$ into \Pzeromode, we can evaluate 
$$
\psi_0(r) \sim \int_\infty^r d{\tilde r}\, {\tilde r}^{-{3 \over 2}} 
	\sim r^{-{1 \over 2}} \,.
$$
So the zero mode $\psi_0$ is normalizable at UV. 
On the other hand, we can estimate $\psi_0$ in the IR region 
by the use of the asymptotic solutions \IRsolution{}, 
$$
\psi_0(r) \sim \int_{r_0}^r d{\tilde r} ({\tilde r} - r_0)^{1-2a}e^{\sqrt{\pi}t_0 \over ({\tilde r}-r_0)^2} \,.
$$
Since the integration kernel exponentially diverges at ${\tilde r}=r_0$, 
the zero mode $\psi_0$ becomes non-normalizable. 
In other words, a massless vector meson does not exist.

\subsec{$A^{(-)}$ part}

The $A^{(-)}$ part of the action \gaugeaction\ is 
\eqn\Aminusact{
S^{{(-)}} = - \int d^4x dr \Bigl[
	\CC_1 \bigl|F^{(-)}_{\mu\nu}\bigr|^2 
	+\CC_2 \bigl|F^{(-)}_{\mu r}\bigr|^2 
	+\CC_3 \bigl|A^{(-)}_\mu\bigr|^2
	+\CC_4 \bigl|A^{(-)}_r \bigr|^2
	+\CC_5 F^{(-)}_{\mu r}A^{(-)\mu}  
\Bigr] \,.
}
We decompose the pseudo-vector field $A^{(-)}_\mu$ 
to the transverse component $A^\perp_\mu$ 
defined by $\partial_\mu A^{\perp \mu} = 0$ 
and the longitudinal one $A^\parallel_\mu$:  
$$
A^{(-)}_\mu = A^\perp_\mu + A^\parallel_\mu \,.
$$
Then we expand these fields by modes,  
\eqna\Aminusmode
$$\eqalignno{
A^\perp_\mu(x^\mu,r) &= \sum_n a^{(-)}_{n\mu} (x^\mu) \xi^\perp_n(r) \,, &\Aminusmode{a} \cr
A^\parallel_\mu(x^\mu,r) &= \sum_n \partial_\mu \omega_n(x^\mu) \xi^\parallel_n(r) \,, &\Aminusmode{b} \cr
A^{(-)}_r(x^\mu,r) &= \sum_n \omega_n(x^\mu) \zeta_n(r) \,. &\Aminusmode{c}
}$$
The modes in \Aminusmode{} are determined by the two eigen equations:
\eqna\Aminuseigeneq
$$\eqalignno{
&-{1 \over 2}{d \over dr}\bigl( \CC_2 \xi^\perp_n{}' \bigr) +{1 \over 2}\CC_3\xi^\perp_n +{1 \over 4}{d\CC_5 \over dr}\xi^\perp_n
	=\bigl(m^{(-)}_n\bigr)^2\CC_1 \xi^\perp_n \,, &\Aminuseigeneq{a} \cr
&\CC_4\zeta_n = M_n^2 \biggl[\CC_2 \bigl( \zeta_n -\xi^\parallel_n{}' \bigr) +{1 \over 2}\CC_5\xi^\parallel_n\biggr] \,, &\Aminuseigeneq{b} 
}$$
and the equation providing the relation between $\zeta_n$ and $\xi^\parallel_n$:
\eqn\Aminusconst{
\partial_r\biggl[\CC_2 \bigl( \zeta_n -\xi^\parallel_n{}' \bigr) +{1 \over 2}\CC_5\xi^\parallel_n\biggr] 
	+\CC_3\xi^\parallel_n +{1 \over 2}\CC_5 \bigl( \zeta_n -\xi^\parallel_n{}' \bigr) = 0 \,,
}
while the normalization conditions are described as 
\eqna\Aminusnorm
$$\eqalignno{
{1 \over 4}\delta_{mn} &= \int dr\, \CC_1\xi^\perp_m\xi^\perp_n \,, &\Aminusnorm{a} \cr
{1 \over 2}\delta_{mn} &= \int dr \biggl[\CC_2 \bigl( \zeta_m -\xi^\parallel_m{}' \bigr) \bigl( \zeta_n -\xi^\parallel_n{}' \bigr) +\CC_3\xi^\parallel_m\xi^\parallel_n \cr
&\qquad	+{1 \over 2}\CC_5 \Bigl( \bigl(\zeta_m -\xi^\parallel_m{}' \bigr)\xi^\parallel_n + \xi^\parallel_m \bigl( \zeta_n -\xi^\parallel_n{}' \bigr) \Bigr)\biggr] \,. &\Aminusnorm{b} 
}$$
In terms of the equations \Aminuseigeneq{}, \Aminusconst\ and \Aminusnorm{}, the action \Aminusact\ is reduced to 
$$
S^{(-)} = -\int d^4x 
\sum_n \biggl[ {1 \over 4}\bigl| f^{(-)}_{n\mu\nu} \bigr|^2 
+{1 \over 2} \bigl(m^{(-)}_n\bigr)^2 \bigl| a^{(-)}_{n\mu} \bigr|^2
+{1 \over 2}\bigl| \partial_\mu \omega_n \bigr|^2 +{1 \over 2} M_n^2 \omega_n^2
\biggr] \,,
$$
where $f^{(-)}_{n\mu\nu} = \partial_\mu a^{(-)}_{n\nu} - \partial_\nu a^{(-)}_{n\mu}$.
$a^{(-)}_{n\mu}$ and $\omega_n$ correspond to the axial-vector meson 
and the pseudo-scalar meson respectively. 

\newsec{Pion}

In this section, we are interested in the lowest mode $\omega_0(x^\mu)$ of pseudo-scalar meson, 
which is identified with a pion. 

\subsec{Pion mass}

Firstly let us assume that the current quark mass $m_q$ vanishes. 
Then the pion should be massless, that is, $M_0=0$, 
by which \Aminuseigeneq{b} leads to 
\eqn\mlzeta{
\zeta_0 = 0 \,.
}
Since the classical solution \UVtacsol\ of tachyon field $T(r)$ 
in the UV region has only normalizable mode 
under that assumption on account of \nnmq, 
$T(r)$ is an exponentially decreasing function. 
This implies that, comparing $\CC_3,\CC_5$ with $\CC_2$, 
we can neglect the terms containing $\CC_3$ or $\CC_5$ in \Aminusconst. 
This approximation allows us to solve \Aminusconst\ in UV, 
\eqn\mlxi{
\xi^\parallel_0 = \alpha + {\beta \over \sqrt{r}} \,,
}
where $\alpha,\beta$ are constants. 
Furthermore the relation between these two constants can be derived 
from the normalization condition \Aminusnorm{b}. 
Substituting \mlzeta\ into \Aminusnorm{b}, we obtain 
\eqn\mlxiperp{
{1 \over 2} = \int_{r_0}^\infty dr \Bigl( \CC_2 \bigl(\xi_0^\parallel{}'\bigr)^2 +\CC_3 \xi_0^{\parallel 2} -\CC_5 \xi_0^\parallel {\xi_0^\parallel}{}' \Bigr) \,.
}
The right hand side of \mlxiperp\ can be evaluated 
by the UV asymptotic solution, because the integrand in the IR region 
does not contribute on account of the tachyon potential. 
Then in terms of \Aminusconst\ incorporated with \mlzeta, 
the integration in \mlxiperp\ is determined by the UV boundary values. 
Finally we obtain the relation between $\alpha$ and $\beta$ 
in the massless quark limit, 
\eqn\albet{
{1 \over 2} = \bigl( \CC_2 \xi^\parallel_0 \xi^\parallel_0{}' \bigr) \big|_{r=\infty} = - {(2\pi\alpha')^2 T_4 \over 2 g_s R^{3/2}} \alpha \beta \,. 
}
However we have not determined $\alpha$ and $\beta$ yet. 
In order to do it, the UV boundary condition of $\xi^\parallel_0$ 
is necessary. Though this condition should be related also to 
the one at IR $(r=r_0)$, we cannot clarify the relation at present. 
Because we know only the UV and IR asymptotic solutions. 
The constants $\alpha,\beta$ will be associated with the pion decay constant 
in the rest of this paper.

Next we shall turn on a small quark mass, 
which allows us to consider the perturbation 
with respect to the small $M_0^2$ and $m_q (\sim C_{\rm nn})$. 
We then set the power series, $\zeta_0 = \zeta_0^{(0)} + M_0^2 \zeta_0^{(1)} + \CO(M_0^4)$ and $\xi^\parallel_0 = \xi_0^{\parallel(0)} + M_0^2 \xi_0^{\parallel(1)} + \CO(M_0^4)$. 
$\zeta_0^{(0)}$ and $\xi_0^{\parallel(0)}$ are identified with \mlzeta\ and \mlxi. 
Since in the UV region we can evaluate $\zeta_0^{(1)}$ from \Aminuseigeneq{b}, 
the UV behavior of $\zeta_0$ becomes 
\eqn\uvzetazero{
\zeta_0 = M_0^2{2\pi\alpha' R^{3 \over 2} \beta \over 8 r^3 T^2} \,.
}
The pion mass square is also described as 
\eqn\pimassint{
M_0^2 = 2\int dr\, \CC_4\zeta_0^2
}
which is derived from the eigen equations \Aminuseigeneq{b} and \Aminusconst\ and the normalization conditions \Aminusnorm{b}. 
Since $\CC_4$ includes the tachyon potential $V(T)$ 
which converges to zero exponentially at the IR limit, 
the IR contribution to \pimassint\ is suppressed. 
Substituting \uvzetazero\ into \pimassint, we obtain 
$$
{1 \over M_0^2} = {R^6 T_4 \Theta_\infty^4\beta^2 \over 16\pi\alpha' g_s}
\int^{r_\infty} {dr \over r}\biggl[C_{\rm n}K_2\biggl({4\pi\alpha'\Theta_\infty \over R^{3/2}}\sqrt{r}\biggr)
	+C_{\rm nn}I_2\biggl({4\pi\alpha'\Theta_\infty \over R^{3/2}}\sqrt{r}\biggr)\biggr]^{-2} \,.
$$
Since $C_{\rm nn}\,(\sim m_q)$ is regarded as being much smaller than $C_{\rm n}$ 
under the small quark mass perturbation, 
the pion mass square is approximately evaluated  
\eqn\pionmass{
M_\pi^2 (=M_0^2) =
	-{8\pi\alpha' g_s C_{\rm n}C_{\rm nn} \over R^6 T_4 \Theta_\infty^4 \beta^2} \,.
}

\subsec{Pion decay constant}

Let us recall that the pion decay constant $f_\pi$ in QCD 
appears in the two-point axial current correlator $\Pi^{(-)}$ 
at the massless quark limit. 
In the large $N_c$ limit, it is described in the momentum space as 
$$
\Pi^{(-)}(p^2) = p^2\sum_n{f_{a^{(-)}_n}^2 \over p^2 + (m^{(-)}_n)^2} + f_\pi^2 \,.
$$
Since, following the AdS/CFT dictionary, the axial-vector current corresponds to 
the UV boundary value of axial-vector field $A^{(-)}_\mu$, 
the effective action of the axial-vector should be 
$$
\int {d^4p \over (2\pi)^2} {1 \over 2}\biggl(\eta^{\mu\nu} - {p^\mu p^\nu \over p^2}\biggr) a^{(-)}_{n\mu} \Pi^{(-)}(p^2) a^{(-)}_{n\nu} \,.
$$ 
Then the pion decay constant can be extracted by differentiating the effective action twice with respect to $a^{(-)}_n$ and putting $p^2 = 0$. 
By the use of this method, Refs.\refs{\DP,\CPS} achieved to 
calculate the pion decay constant in AdS/QCD model and Ref.\BSS\ in the modified SS model. 

We shall compute the pion decay constant in the same way. 
We concentrate on the zero mode of the axial-vector field 
$a^{(-)}_{0\mu}\xi^\perp_0$ and consider its Fourier transformation 
$a^{(-)}_{0\mu}(x^\mu) = \int {d^4p \over (2\pi)^2}e^{ipx}a^{(-)}_{0\mu}(p^\mu)$. 
Then we extract from \Aminusact\ the effective action of $a^{(-)}_{0\mu}$ 
on the momentum space, 
$$\eqalign{
S^{(-)}_\perp = -\int {d^4p \over (2\pi)^2} \int_{r_0}^\infty dr \Bigl[
	&\bigl( 2\CC_1 \xi_0^{\perp 2} p^2 + \CC_2 \bigl(\xi_0^\perp{}' \bigr)^2 +\CC_3 \xi_0^{\perp 2} -\CC_5 \xi_0^\perp \xi_0^\perp{}' \bigr)a^{(-)}_{0\mu} a^{(-)\mu}_0 \cr
	&-2\CC_1 \xi_0^{\perp 2} (p_\mu a^{(-)\mu}_0)^2
\Bigr] \,. 
}$$
From this equation, we can read the pion decay constant,
\eqnn\fpipre
$$\eqalignno{
{1 \over 2}f_\pi^2 &= \int_{r_0}^\infty dr \Bigl( \CC_2 \bigl(\xi_0^\perp{}'\bigr)^2 +\CC_3 \xi_0^{\perp 2} -\CC_5 \xi_0^\perp \xi_0^\perp{}' \Bigr)  &\fpipre \cr
&= \int_{r_0}^\infty dr {d \over dr} \biggl(\CC_2 \xi_0^\perp \xi_0^\perp{}' -{1 \over 2}\CC_5 \xi_0^{\perp 2} \biggr) \,, 
}$$ 
where we used the eigen equations. 
The IR region does not contribute to the integration in this equation, 
because $\CC_2, \CC_5$ include tachyon potential $V(T)$ which 
exponentially vanishes by $r\to r_0$. 
On the other hand, in the UV region, 
$\CC_2$ is dominant compared with $\CC_5$ and provide the leading contribution for the integration in \fpipre.   
Finally we evaluate \fpipre\ as
\eqn\fpibdry{
{1 \over 2}f_\pi^2 
= \bigl(\CC_2 \xi_0^\perp \xi_0^\perp{}' \bigr)\big|_{r = \infty} \,, 
}
where we used \Aminuseigeneq{a} with $m^{(-)}_0 = 0$, in other words, 
the on-shell condition for $A^{\perp}_\mu$. 
\fpibdry\ implies that the pion decay constant is described 
in terms of the UV boundary value. 

Comparing \fpipre\ with \mlxiperp, we can read the relation,  
\eqn\xiparaperp{
{\xi^\perp_0 \over f_\pi} = \xi^\parallel_0 \,. 
}
If we fix the boundary condition $\xi_0^\perp(\infty) = c$, where 
$c$ is a dimensionless constant, 
\mlxi, \fpibdry\ and \xiparaperp\ allow us to describe $\alpha,\beta$ by the use of $f_\pi$ as 
\eqn\albetfpi{
\alpha = {c \over f_\pi} \,,\quad 
\beta= -{g_s R^{3 \over 2} f_\pi \over (2\pi\alpha')^2 T_4 c} \,. 
}

\subsec{Gell-Mann-Oakes-Renner relation}

We have calculated the quark mass \nnmq, the quark condensate \nqcond\ 
and the pion mass \pionmass. 
Combining these quantities with \albetfpi, 
we can show GOR relation, 
\eqn\gorrel{
M_\pi^2 = -8 c^2 {m_q \langle q{\bar q}\rangle \over f_\pi^2} \,, 
}
up to a numerical factor $4c^2$. 
In order to satisfy an exact GOR relation, 
the factor $c$, which is the UV boundary value $\xi^\perp_0(\infty)$, has to be 
equal to $1/2$. However, as we mentioned, we cannot determine this value, because we do not know the exact classical solutions over all $r\,(\geq r_0)$. 

\newsec{Scalar fields}

In this section, we shall give some comments on scalar fields. 
So far we have studied the gauge fields, 
which contain a pseudo-scalar field $A^{(-)}_r$. 
There still exist many fluctuations of scalar fields 
which originate from the collective coordinates $x_4, \theta, {\vec x}_T$ 
and the tachyon $T$. 

For instance, let us turn on the fluctuation along the $x_4$ direction. 
In VW model, 
the quadratic effective Lagrangian for the fluctuation of $x_4$ around the classical solution does not 
include its mass term, so that this fluctuation provides us with a massless mode 
in the non-compact limit ($\rho_{\rm KK}\to 0$).
On the other hand, in our intersecting D4-branes model, 
the fluctuations of $x_4$ obtain a mass term in the way similar to the gauge fields. 
Concretely we set the fluctuations as $X_4^{(\pm)}(x^\mu,r) = x_4^{(1)} \pm x_4^{(2)}$. Note that $X_4^{(-)}(r_0) = X_4^{(-)}{}'(r_0) = 0$ is necessary for the smoothness at the tip of the D4-branes. 
Then we calculate the quadratic effective action, 
$$\eqalign{
S[X_4] &= -{T_4 \over g_s}\int d^4x dr \biggl[
{V(T) \over 4}\sqrt{\CD}\biggl({r \over R}\biggr)^{3 \over 2} \Bigl( \bigl|\partial_\mu X_4^{(+)}\bigr|^2 +\bigl|\partial_\mu X_4^{(-)}\bigr|^2 \Bigr) \cr
&\quad	+{V(T) \over 4\sqrt{\CD}}\biggl({r \over R}\biggr)^{9 \over 2}\biggl(1+{r^2 \over 2\pi\alpha'}\biggl({R \over r}\biggr)^{3 \over 2} \Theta^2 T^2 \biggr) \Bigl(\bigl(X_4^{(+)}{}'\bigr)^2 +\bigl(X_4^{(-)}{}'\bigr)^2 \Bigr) \cr
&\quad	+{V(T) \over 2\pi\alpha'\sqrt{\CD}}\biggl({r \over R}\biggr)^3 \biggl(1+{\Theta'^2 \over 4} \biggr)T^2 \bigl(X_4^{(-)}\bigr)^2 
-{V(T) r^2 \over 8\pi\alpha' \sqrt{\CD}}\biggl({r \over R}\biggr)^3\Theta\Theta'T^2 X_4^{(-)} X_4^{(+)}{}'  \cr
&\quad	-{r^3 \over R^3} X_4^{(-)}{}' \biggr]\,,
}$$
where $\Theta(r)$ and $T(r)$ are the classical solutions. 
The last term is derived from the Chern-Simon action. 
As we mentioned above, the mass term of $X^{(-)}_4$ 
which is proportional to $T^2$ appears explicitly.

\newsec{Conclusions and discussions}

We have studied the intersecting D4-branes  
in the background of large $N_c$ D4-branes by the use of the tachyonic DBI action. 
We have found the trivial solution of the equations of motion 
which corresponds to the intersecting D4-branes with $T(r)=0$. 
In this solution the tachyon stays on the top of the potential $V(T)$. 
Since the bi-fundamental ``tachyon'' field, 
which originates from the open string stretched between the intersecting D4-branes, 
has negative mass square around the intersection point, 
the D4-branes recombine into the U-shaped D4-branes. 
We have analyzed the asymptotic behavior 
by analytically solving the equations of motion in the IR and UV regions, 
and we have obtained the U-shaped classical solution corresponding to that recombination. 
We have also computed the solutions numerically from the IR or UV initial conditions. 
However at present it is difficult to find the full solutions 
in all region even numerically. This issue is left for future works. 

The classical solution of the tachyon field in UV consists of the 
non-normalizable mode and the normalizable one. 
By assuming that the former corresponds to the current quark mass, 
we have shown that the latter is naturally related to the quark condensate.

The effective action for the fluctuations of the gauge fields from the classical solution has been calculated 
and contains the mass terms which appear 
due to the tachyon field. 
By the mode expansions of these fluctuations, 
the vector, axial-vector and pseudo-scalar mesons 
have been obtained in the $A_r^{(+)} = 0$ gauge. 
We have shown that the massless vector meson does not exist. 
And we have also evaluated the pion mass and the pion decay constant 
by the use of the UV and IR asymptotic classical solutions. 
Incorporating the current quark mass and the quark condensate 
into these quantities of the pion, 
we have obtained GOR relation up to a numerical factor. 

In our model, there are also the scalar fluctuations derived from the collective coordinates of the probe D4-branes. 
We have computed the fluctuation of $x_4$ direction, so that it has a mass term including the tachyon field in the similar way to the gauge fields. 
On the other hand, without the tachyon, 
our model is reduced to the model of \VW, 
in which there is no mass term for the $x_4$ fluctuation. 

In the rest of this paper, we shall give some comments. 
Though we have considered the separation of D4-branes along only the $\theta$ direction, 
the separations along the $x_4$ and ${\vec x}_T$ directions are also available. 
In this case we must consider the contribution 
by the Chern-Simons action \dadCS.
For instance, let us additionally turn on the separation $L(r)$ 
along the $x_4$ directions defined 
by $x_4^{\rm D4} = -x_4^{\rm \overline{D4}} = L(r)/2$ (see fig.~5). 
The action is written down 
as $S[\Theta,T,L] = S_{\rm DBI}[\Theta,T,L] + S_{\rm CS}[L]$, 
$$\eqalign{
S_{\rm DBI} &= -{2T_4 \over g_s} \int d^4x dr \, V(T)
	\biggl({r \over R}\biggr)^{3 \over 2} \sqrt{\CD_L} \,, \cr
&\CD_L = 1 +{1 \over 4}\biggl(r^2\Theta'^2 +\biggl({r \over R}\biggr)^3L'^2\biggr) +2\pi\alpha'\biggl({r \over R}\biggr)^{3 \over 2}T'^2 \cr
&\qquad	+{1 \over 2\pi\alpha'}\biggl({R \over r}\biggr)^{3 \over 2}\biggl(r^2\Theta^2 
	+\biggl({r \over R}\biggr)^3 L^2 \biggr)T^2
		+{r^2 \over 8\pi\alpha'}\biggl({r \over R}\biggr)^{3 \over 2}(\Theta L' -L \Theta')^2 T^2 \,, \cr
S_{\rm CS} &= {N_c \over 16\pi^3\alpha' R^6}\int d^4xdr\, r^3 \biggl({d x_4^{\rm D4}\over dr} - {d x_4^{\rm \overline{D4}}\over dr}\biggr) 
	= {N_c \over 16\pi^3\alpha' R^6}\int d^4xdr\, r^3 L'
}$$
\bigskip 
\vbox{\centerline{\epsfbox{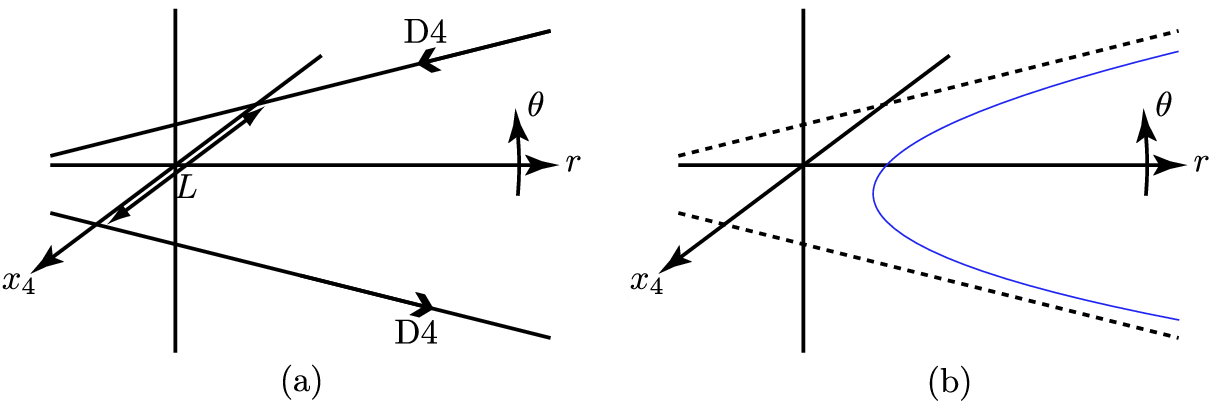}} 
\centerline{\fig\figshearU{} (a) The skew D4-branes. (b) The sheared U-shaped D4-brane.}} 
\bigskip\nobreak\noindent 
If $L(0)$ is sufficiently small, 
the ``tachyon'' field $T$ becomes really tachyonic around $r=0$ 
by the same reason discussed in the previous sections. 
The skew D4-brane and anti-D4-brane therefore are connected 
with each other through the tachyon condensation, 
and roughly lead to the sheared U-shaped D4-branes like fig.~5(b). 
However, as discussed in \VW, 
the shape of the flavor branes would be modified 
owing to the Chern-Simons action. 
This issue is left for future work. 
We should note that the new parameters 
which are derived from the boundary conditions for $L(r)$ 
can be introduced additionally into the dual gauge theory.

So far we have considered the non-compact background, 
because it is impossible to put such intersecting D4-branes 
in the compact background ($\rho_{\rm KK} \neq 0$). 
However the UV behavior would not be changed even in the compact case. 
If we formally write down the action in the compact background as 
$$\eqalign{
&S_{\rm cpt} = -{2T_4 \over g_s}\int d^4x dr \, V(T)
	\biggl({r \over R}\biggr)^{3 \over 2} g_+^{5 \over 3} 
	\sqrt{\CD_{\rm cpt}} \,, \cr
&\quad	\CD_{\rm cpt} = 1+{r^2 \over 4}\Theta'^2 
	+{2\pi\alpha' \over g_+^{1/3}}\biggl({r \over R}\biggr)^{3 \over 2}T'^2
	+{g_+^{1/3} r^2 \over 2\pi\alpha'}\biggl({R \over r}\biggr)^{3 \over 2} \phi^2 T^2 \,, 
}$$
the difference between the non-compact and the compact cases is the factor $g_+(r)$. 
Since it is equal to one at the limit $r\to \infty$, 
the difference disappears in the UV region. 
Furthermore GOR relation would be satisfied also in the compact case, 
because the quantities about the quark and the pion which we have calculated 
are described in terms of the UV boundary values. 

 


\appendix{A}{Parallel D4-brane and anti-D4-brane}

We can describe the $\Bbb{R}^5$ part in \ncmet\ in terms of the Cartesian coordinates $(y,z,{\vec x}_T)$ as 
$$
ds^2 = \biggl({\rho \over R}\biggr)^{3 \over 2} 
	\bigl[\eta_{\mu\nu}dx^\mu dx^\nu +dx_4^2 \bigr] 
	+\biggl({R \over \rho}\biggr)^{3 \over 2}
	\bigl[ dy^2 + dz^2 +d{\vec x}_T^2 \bigr] \,, 
$$
where $\rho^2 = y^2 +z^2 +{\vec x}_T^2$.
In the analogy of \BSS, it seems possible to locate 
the parallel D4-branes and anti-D4-branes 
which are separated along the $z$ direction. 
When the D4-brane and anti-D4-brane with the world-volume 
$(x^0,x^1,x^2,x^3,y)$ are embedded into the target space 
so that $x_4^{\rm D4} = x_4^{\rm \overline{D4}}=0$, 
$z^{\rm D4} = -z^{\rm \overline{D4}}=Z(y)$ 
and ${\vec x}_T^{\rm D4} = {\vec x}_T^{\rm \overline{D4}} = 0$, 
the action \dadaction\ becomes 
$$\eqalign{
&S[Z,T] = -{2T_4 \over g_s}\int d^4x dy \, V(T)
	\biggl({\rho \over R}\biggr)^{3 \over 2}  
	\sqrt{\CD_Z} \,, \cr
&\quad \CD_Z = 1+{1 \over 4}\biggl({dZ \over dy}\biggr)^2 
	+2\pi\alpha' \biggl({\rho \over R}\biggr)^{3 \over 2}\biggl({dT \over dy}\biggr)^2
	+{1 \over 2\pi\alpha'}\biggl({R \over \rho}\biggr)^{3 \over 2}Z^2 T^2 \,, \quad \rho^2 = {1 \over 4}Z^2 + y^2 \,. 
}$$
From this action, we calculate the equations of motion 
for $Z(y)$ and $T(y)$:
\eqna\paraeom
$$\eqalignno{
&{d \over dy}\biggl[{V(T) \over \sqrt{\CD_Z}}\biggl({\rho \over R}\biggr)^{3 \over 2}{dZ \over dy}\biggr] \cr
&\quad	= {3 \over 2}V(T) {Z \sqrt{\CD_Z} \over R^{3 \over 2}\rho^{1 \over 2}} 
+ {V(T) \over \sqrt{\CD_Z}} \biggl[
	{3 \over 4}{2\pi\alpha' \rho Z \over R^3}\biggl({dT \over dy}\biggr)^2
	-{3 \over 4}{Z^3 T^2 \over 2\pi\alpha' \rho^2}
	+{4 Z T^2 \over 2\pi\alpha'} \biggr] \,, &\paraeom{a} \cr
&{d \over dy}\biggl[ {2\pi\alpha' V(T) \over \sqrt{\CD_Z}}\biggl({\rho \over R}\biggr)^3  {dT \over dy} \biggr] 
= {V(T) \over 2\pi\alpha' \sqrt{\CD_Z}} Z^2 T 
	+\biggl({\rho \over R}\biggr)^{3 \over 2} \sqrt{\CD_Z} {dV(T) \over dT} \,. &\paraeom{b}
}$$
The separated parallel D4-brane and anti-D4-brane which are denoted 
by $Z(y)= {\rm constant} \neq 0$ and $T(y)=0$ are not a solution. 
On the other hand, $Z(y)=0$ and $T(y)=0$ are the solution of \paraeom{}, 
however it describes the coincident D4-brane and anti-D4-brane, 
which are the same configuration given by the solution $\Theta = 0$ 
and $T=0$ of the intersecting case \itseom{}.

\listrefs
\bye